\documentclass[a4paper]{jpconf}
\usepackage{graphicx}
\usepackage{here}

\def\frac#1#2{\textstyle{{{#1} \over {#2}}}}

\def\lsim{\mathrel{\rlap{\lower4pt\hbox{\hskip1pt$\sim$}}
    \raise1pt\hbox{$<$}}}
\def\gsim{\mathrel{\rlap{\lower4pt\hbox{\hskip1pt$\sim$}}
    \raise1pt\hbox{$>$}}}

\newcommand{\beq}{\begin{equation}}
\newcommand{\eeq}{\end{equation}}
\newcommand{\bea}{\begin{eqnarray}}
\newcommand{\eea}{\end{eqnarray}}

\begin{document}

\title{Stringy Space-Time Foam and High-Energy Cosmic Photons}

\author{Nick E. Mavromatos}

\address{CERN, Theory Division, CH-1211 Geneva 23, Switzerland; \\ On leave from : King's College London, Physics Department, Strand, London WC2R 2LS, UK}

\ead{nikolaos.mavromatos@kcl.ac.uk}

\begin{abstract}

In this review, I discuss briefly stringent tests of Lorentz-violating quantum space-time foam models inspired from String/Brane theories, provided by studies of high energy Photons from intense celestial sources, such as Active Galactic Nuclei or Gamma Ray Bursts. The theoretical models predict modifications to the radiation dispersion relations, which are quadratically suppressed by the string mass scale, and  time delays in the arrival times of photons (assumed to be emitted more or less simultaneously from the source), which are proportional to the photon energy, so that the more energetic photons arrive later. Although the astrophysics at the source of these energetic photons is still not understood, and such non simultaneous arrivals, that have been observed recently, might well be due to non simultaneous emission as a result of conventional physics effects, nevertheless, rather surprisingly, the observed time delays can also fit excellently the stringy space-time foam scenarios, provided the space-time defect foam is inhomogeneous.
The key features of the model, that allow it to evade a plethora of astrophysical constraints on Lorentz violation, in sharp contrast to other field-theoretic Lorentz-violating models of quantum gravity,
are: (i) transparency of the foam to electrons and in general charged matter, (ii) absence of birefringence effects and (iii) a breakdown of the local effective lagrangian formalism.
\end{abstract}

\section{Introduction: Mysterious Results in High Energy  Gamma-Ray Astronomy}
\vspace{0.1cm}
\paragraph{}
On July 9th 2005, the MAGIC (\emph{\textbf{M}}ajor \emph{\textbf{A}}tmospheric \emph{\textbf{G}}amma-ray \emph{\textbf{I}}maging \emph{\textbf{C}}herenkov) Telescope, located in the Canary Islands, observed~\cite{MAGIC} very high energy gamma rays from the active galactic nucleus Markarian 501 (Mkn 501),
which lies at red-shift $z=0.034$  (\emph{i.e.} about half a million light years away) from Earth,
with energies up to the order of 10 TeV  (1 TeV = $10^3$ GeV = $10^{12}$ eV), which were delayed up to four minutes as compared with their lower-energy counterparts (in the 0.6 TeV or lower range). It was the first observation
of such a distinct delay.

Three years after the MAGIC observations, in September 2008, the FERMI (formerly known as GLAST) Satellite Telescope~\cite{glast}, also observed time delays of the higher-energy photons, from the distant Gamma Ray Burst (GRB) 080916c~\cite{grbglast}, at red-shifts $z = 4.35$,  and later on from GRB 090510~\cite{grb090510}, at red-shift $z=0.9$ and from GRB 09092B, at redhifts $z = 1.822$~\cite{grb09092b}.

The delay effects may be due to the conventional astrophysics of the active galactic nucleus or Gamma-Ray Burst (\emph{source} effect), which, however, is not well understood at present~\cite{deangelis,mavro_review}. In fact, currently there seem to be no consensus among the astrophysicists on the appropriate mechanism for the production of such high-energy photons at the source.

These uncertainties prompted more ambitious, although admittedly far-fetched, explanations for the MAGIC effect~\cite{MAGIC2} and more general for the other observed delayed arrivals of photons,
which pertain to  new fundamental physics, affecting the photon \emph{propagation}. This may be due, for instance, to space-time \emph{foamy} vacuum structures due to quantum gravitational fluctuations~\cite{wheeler} that lead to modified  dispersion relations for photons~\cite{mitsou,robust,JP}. If true, this would be a clear
departure from the Lorentz invariant energy (E)-momentum ($\vec p$) relations of Special Relativity, $E=|\vec p|c$.

\begin{figure}[H]
\centering
\includegraphics[width=7.5cm,angle=-90]{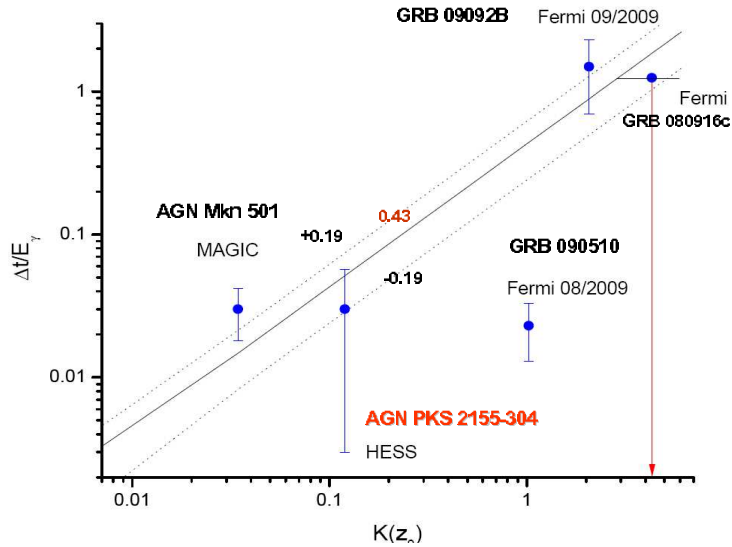}
\caption{Comparison of data on delays
$\Delta t$ in the the arrival times of energetic gamma rays from various astrophysical sources with
models in which the velocity of light is reduced by an amount linear in the photon
energy. The graph plots on a logarithmic scale the quantity $\Delta t/E$ and a function of the
red-shift, $K(z)$, which is essentially the distance of the source from the observation point.
The data include two Active Galactic Nuclei (AGN), Mkn 501~\protect\cite{MAGIC2} and PKS
2155-304~\protect\cite{hessnew}, and three Gamma Ray Bursts (GRB) observed by the Fermi satellite,
090510~\cite{grb090510}, 09092B~\cite{grb09092b} and 080916c~\cite{grbglast}.}%
\label{fig:data}
\end{figure}
If we plot the observed time delays $\Delta t$ versus the energy $E$ of the photon, including experimental errors, we obtain~\cite{emnfit} the diagram of fig.~\ref{fig:data}. We observe from the figure that
with the exception of GRB 090510 , observed by the FERMI satellite~\cite{grb090510}, the other four points can be fitted by a straight line:
\begin{equation}
\Delta t/E_\gamma = (0.43 \pm 0.19) \times K(z) {\rm s/GeV} ,~\quad
K(z) \equiv   \int_0^z \frac{(1 + z)dz}{\sqrt{\Omega_\Lambda + \Omega_m (1 + z)^3}} ,
\label{bestfit}
\end{equation}
assuming an expanding Universe within the framework of the standard Cosmological-constant-Cold-Dark-Matter ($\Lambda$CDM) model.
The function $K(z)$ expresses the effect of an expanding Universe onto the linear relation $\Delta t \propto E$, and includes (i) a time dilation factor~\cite{JP} $(1 + z)$ and (ii) the redshifting~\cite{robust}
of the photon energy which implies that the observed energy of a photon
with initial energy $E$ is reduced to $E_{\rm obs} = E_0/(1 + z)$. Taking into account that in a
Robertson-Walker expanding Universe, we assume for our analysis, the
infinitesimal time interval $dt$ is related to the Hubble rate $H(z)$, where $z$ is the red-shift, via:
$dt = - dz/[(1 + z) H(z)]$, we obtain a total delay in the arrival times of photons with energy difference $\Delta E$:
\begin{equation}
(\Delta t)_{\rm obs} =  \frac{ \Delta E}{M_{\rm QG1}} {\rm H}_0^{-1}\int_0^z  \frac{(1 + z)dz}{\sqrt{\Omega_\Lambda + \Omega_m (1 + z)^3}}
\label{redshift}
\end{equation}
where ${\rm H}_0$ is the (current-era) Hubble expansion rate, which is fitted to the data, (\ref{bestfit}).
As already mentioned, we have assumed for concreteness the $\Lambda$CDM standard model of cosmology, with
$\Omega_i \equiv \frac{\rho_{i(0)}}{\rho_c}$ representing the present-epoch energy densities ($\rho_i$) of matter (including dark matter), $\Omega_m$, and dark vacuum energy, $\Omega_\Lambda$, in units of the critical density $\rho_c \equiv \frac{3H_0^2}{8\pi G_N}$ of the Universe ($G_N$ is the Newton's gravitational constant). The current astrophysical measurements of the acceleration of the Universe are all consistent with a non zero Cosmological-Constant Universe with Cold-Dark-Matter ($\Lambda$CDM Model), with $\Omega_\Lambda \sim 73 \%$ and $\Omega_m \sim 27\% $. The quantity $M_{\rm QG1}$ appearing in (\ref{redshift}) may be viewed as a phenomenological parameter at this stage, with units of energy. In concrete models of space-time foam, where propagation effects due to QG are assumed to be the dominant cause of the delay, this scale will be identified with the Quantum Gravity scale, where space-time foam effects are expected to set in.

The best fit~\cite{emnfit}, based on (\ref{bestfit}), leads to the following result for the
scale, $M_{QG1} = (0.98^{+0.77}_{-0.30}) \times 10^{18}$~GeV. This result is remarkable, because it provides the first hint that such delays might be related with string theory effects, given that the order of the scale is that of the conventional string mass scale~\cite{polch}.

However, the GRB 090510 data, provided that the latter can be trusted~\footnote{Indeed, there are uncertainties in this measurement concerning, for instance, the precise emission time of photons due to the yet uncertain duration of pre-cursors to burst activities and other such issues, which need to be confirmed by other measurements of similarly short bursts, that are presently lacking.}, do not fit to this linear energy scheme. In fact the time delays and energies pertaining to the GRB 090510~\cite{grb090510} are such that the pertinent energy scale $M_{\rm QG1}$ in (\ref{bestfit}), that would reproduce this point in the graph of fig.~\ref{fig:data}, is found to be at least $1.2 M_P$, with $M_P \sim 1.2. \times 10^{19}~{\rm GeV}$ the Planck mass.

The immediate reaction to this result, of course, is~\cite{grb090510} to \emph{exclude} the QG as being responsible for the induced time delays, since one expects, on accounts of naturalness, that any QG effect should not appear at scales higher than the Planck scale. And, indeed, this may well be the case, since, as we have already mentioned, astrophysical mechanisms at the source, which are in general different between GRBs and AGNs, may be responsible for the observed delayed arrivals of the more energetic photons.
Nevertheless, as we shall discuss below, the pattern of the observed photon delays fits~\cite{emnfit} a string model of quantum-gravity-induced refractive index, with the pertinent quantum gravity energy scale being essentially the same as that inferred from the MAGIC observations (of order $10^{18}$~GeV). Viewed as a lower bound, this scale is also compatible with that obtained from other Gamma-Ray data of the H.E.S.S. Collaboration~\cite{hess2155,hessnew}.
In particular, we shall argue that a stringy model of space-time foam (``D-foam''), proposed in \cite{emnw,emnnewuncert,li}, which is based on scenarios involving brane worlds punctured by localized space-time brany defects (termed D-``particles''), can provide consistent fits to the current data on delayed arrival times of photons
from MAGIC and FERMI Telescopes.

However, in order to accommodate the findings of MAGIC with the recently observed time delays of photons from the extremely short Gamma-Ray burst GRB 090510~\cite{grb090510}, one needs inhomogeneous densities of D-particle defects in the foam~\cite{emndvoid}. In this sense, astrophysical observations of such intense, short Bursts, are quite essential in falsifying models. The important feature of the string model, which allows for a consistent fit to all available data, including the GRB 090510, is that the foam there consists of real and not virtual space-time defects. Hence, their density may be \emph{non-uniform}, varying with the redshift, \emph{i.e}. be different at various epochs of the Universe. As we shall discuss below~\cite{emndvoid}, the effective QG scale that enters (\ref{redshift}) is inversely proportional to this density. A depletion or reduction, therefore, of the number of defects encountered by the photons at the redshift region of the GRB 090510, $z=0.9$ (which by the way is the regime where one expects a cosmic deceleration to acceleration transition for the Universe) may well be responsible for a reconciliation of the observed delays for the GRB 090510
with those of MAGIC, in the sense that they are both due to the (inhomogeneous) stringy foam.
 From this point of view, the very few available data on observed time delays available today (fig.~\ref{fig:data}) are not sufficient to exclude the model. One needs a significant improvement on statistics of relevant measurements at various redshifts and various directions (in order to exclude anisotropic foam situations), before definite conclusions on the falsification of the model are reached. Complementary tests from the Cosmology of these models may be helpful, although for inhomogeneous foam situations, studies in the early universe epoch cannot be used to constrain the present-era density of foam, as the two could be different.

\section{Lorentz-Violating Stringy Space-time Foam model: D-foam\label{sec:dfoam}}
\vspace{0.1cm}
\paragraph{}
One of the cornerstones of Modern Physics is Einstein's theory of Special Relativity (SR), which is based on the assumption that the speed of light in vacuo $c$ is an invariant under all observers. In fact, this implies
invariance of the physical laws under the Lorentz transformations in flat space times, and the r\^ole of $c$ as a universal limiting velocity for \emph{all} particle species.
The generalization (by Einstein) of SR to include curved space times, that is the theory of General Relativity (GR), encompasses SR locally in the sense of the \emph{strong form of the equivalence principle}. According to it,  \emph{at every space-time point, in an arbitrary gravitational
field, it is possible to choose a locally
inertial (`free-float')  coordinate frame, such that within a sufficiently
small region of space and time around the point in question,
the laws of Nature are described by special relativity, \emph{i.e.} are of the
same form as in unaccelerated Cartesian coordinate frames in the absence
of Gravitation.} In other words, locally one can always make a coordinate transformation
such that the space time looks {\it flat}. This is not true globally, of course, and this is why GR is a more general theory to describe gravitation. The equivalence principle relies on another fundamental invariance of GR, that of general coordinate, that is the invariance of the gravitational action under arbitrary changes of coordinates. This allows GR to be expressed in a generally covariant form.

In such a locally \emph{Lorentz-invariant} vacuum,
the photon dispersion relation, that is a local in space-time relation between the photon's four-wavevector components $k^\mu =(\omega, \vec k)$ (where $\omega$ denotes the frequency, and $\vec k$ the momentum) reads in a covariant notation:
\begin{equation}
  k^\mu k^\nu \eta_{\mu\nu} = 0
\label{photonflat}
\end{equation}
where repeated indices $\mu, \nu = 0, 1, \dots 3$, with $0$ referring to temporal components, denote summation and $\eta_{\mu\nu}$ denotes the Minkowski space-time metric, with components $\eta_{00} = -1, ~\eta_{0i}=\eta_{i0}=0, ~ \eta_{ij}=\eta_{ji}=\delta_{ij}~, i = 1,2,3$ with $\delta_{ij}$ the Kronecker delta symbol.

The above relation (\ref{photonflat}) implies the equality of all three kinds of photon velocities \emph{in vacuo} that stem from its wave nature (due to the particle-wave duality relation):
\begin{eqnarray}
&& {\rm phase}: \qquad v_{\rm ph} = \frac{\omega}{|\vec k|} \equiv \frac{c}{n(\omega)} = c \nonumber \\
&& {\rm group}: \qquad v_{\rm gr} = \frac{\partial \omega }{\partial |\vec k|} \equiv \frac{c}{n_{\rm gr}(\omega)} = c~,~ \quad n_{\rm gr}(\omega) = n(\omega) + \omega \frac{\partial n(\omega)}{\partial \omega}  \nonumber \\
&& {\rm front}: \qquad v_{\rm front} = c/n(\infty) = c
\label{velocities}
\end{eqnarray}
since the phase and group \emph{refractive indices} of the trivial \emph{vacuum } equal unity $n(\omega) = n_{\rm gr}(\omega) = 1$. For brevity we shall work from now on in units where $\hbar = c=1$.

GR is a classical field theory. In a quantum theory of space-time, i.e in a Quantum-Gravity (QG) model, there is a priori no fundamental reason for the strong equivalence principle to hold, and in this sense, local Lorentz invariance might not be respected. This is the situation, in fact, that characterizes certain QG models, which predict a space-time ``\emph{foamy}'' structure at microscopic (Planck) scales~\cite{wheeler}. The latter may not be Lorentz invariant.
The foamy structures could include a variety of topologically non-trivial configurations at such microscopic scales, ranging from virtual black holes to space-time \emph{non-commutativity} and (real) stringy defects, as is the case of our stringy D-particle foam models~\cite{emnw,emnnewuncert,li}, which we now proceed to review.

\begin{figure}[H]\begin{center}
  \includegraphics[width=8cm]{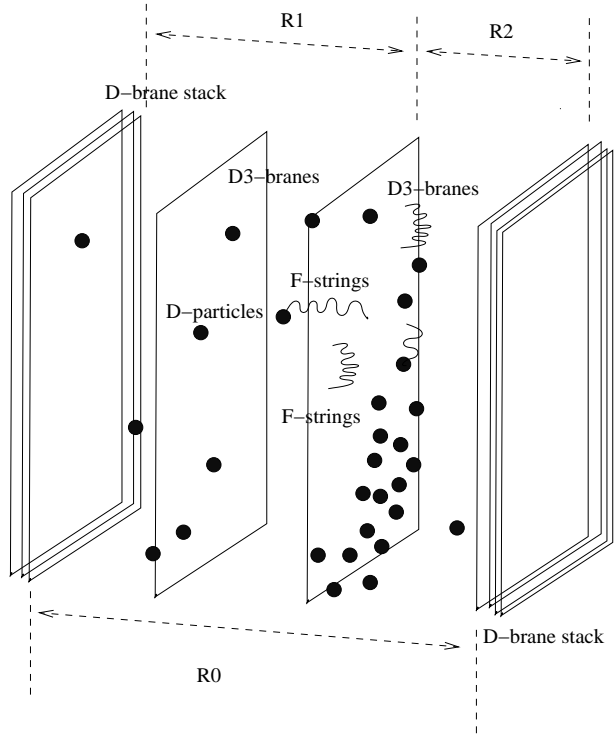}
\end{center}\caption{A string theory model of D-particle ``foam''. The model consists of appropriate stucks of parallel D(irichlet)branes, some of which are moving in a higher-dimensional Bulk space time, punctured by point-like D-brane defects (D-particles). The observable Universe is represented by one of such moving branes, compactified to three spatial dimensions (D3-brane). As the D3-brane world moves, D-particles from the bulk cross it and, thus, to an observer on the brane, they appear as ``flashing on and off'' space-time foam defects (``D-particle foam''). Photons, represented by open strings with their ends attached on the D3 brane, interact with these defects via capture/recoil, and this leads to non-trivial refractive indices. The effect is therefore ``classical'' from the bulk space time point of view, but appears as an effective ``quantum foam'' from the D3-brane observer effective viewpoint.}
\label{dfoam:fig}
\end{figure}

The model of \cite{emnw}
constitutes an attempt to construct a brane/string-inspired
model of space time foam which could have realistic cosmological properties.
For this purpose we exploited the modern approach to string theory~\cite{polch}, involving
membrane hypersurfaces (D(irichlet)-branes). Such structures are responsible for reconciliating (often via duality symmetries) certain string theories (like type I), which before were discarded as physically uninteresting, with Standard-Model phenomenology in the low-energy limit.

In particular, we
considered (\emph{c.f}. figure \ref{dfoam:fig}) a ten-dimensional bulk bounded
by two eight-dimensional orientifold planes, which contains two stacks of
eight-dimensional branes, compactified to three spatial
dimensions. Owing to special reflective properties, the orbifolds act as
boundaries of the ninth-dimension. The bulk space is punctured by point-like D0-branes
(D-particles), which are allowed in type IA string theory (a T-dual of type I strings~\cite{schwarz}) we consider in \cite{emnw} and here~\footnote{One can extend the construction to phenomenologically realistic models of type IIB strings~\cite{li}.}. These are massive objects in string theory~\cite{polch}, with masses $M_s/g_s$, where $M_s$ is the string mass scale (playing the r\^ole of the quantum gravity scale in string theory), and $g_s < 1$ is the string coupling, assumed weak for our purposes. These
objects are viewed as space-time \emph{defects}, analogous, \emph{e.g.} to cosmic strings, but these are point-like and electrically neutral. I have to stress at this point that, according to modern ideas in string theory~\cite{polch}, the scale $M_s$ is in general different from the four-dimensional Planck-mass scale $M_P = 1.2 \times 10^{19}$ GeV/$c^2$, and in fact it is a free parameter in string theory to be constrained by experiment.
The energy scale $M_s c^2$ can be as low as a few TeV; it cannot be lower than this, though, since if this were the  case we should have already seen fundamental string structures experimentally.

Supersymmetry dictates the number of
D8-branes in each stack in the model, but does not restrict the number of D0-branes in the bulk.
For definiteness, we restrict our attention for now to the type-IA model, in which the
bulk space is restricted to a finite range by two appropriate stacks of D8-branes, each
stack being supplemented by an appropriate orientifold eight-plane with specific
reflecting properties, so that the bulk space-time is effectively compactified to a finite
region, as illustrated in Fig.~\ref{dfoam:fig}. We then postulate that two of the D8-branes
have been detached from their respective stacks, and are propagating in the bulk.
As discussed above, the bulk region is punctured by D0-branes (D-particles),
whose density may be \emph{inhomogeneous}.
When there are no relative motions of the D3-branes or D-particles,
it was shown in~\cite{emnw} that the ground-state energy vanishes, as decreed by
the supersymmetries of the configuration. Thus, such static configurations constitute appropriate
ground states of string/brane theory. On the other hand, relative motions of the branes break
target-space supersymmetry explicitly, contributing to the dark energy density.

Our model assumes a collision between two branes from the original stack of branes (\emph{c.f.} figure \ref{dfoam:fig}) at an early
epoch of the Universe, resulting in an initial cosmically catastrophic
Big-Bang type event in such non-equilibrium cosmologies~\cite{brany}. After
the collision, the branes bounce back.
It is natural to assume that, during the current (late) era of the Universe, the D3-brane
representing our Universe is moving slowly
towards the stack of branes from which they
emanated and the configuration is evolving
adiabatically. Hence populations of bulk D-particles cross the brane worlds and
interact with the stringy matter on them. To an observer on the brane the
space-time defects will appear to be {\em \textquotedblleft
flashing\textquotedblright } \emph{on} and \emph{off}. The model we are using involves
eight-dimensional branes and so requires an appropriate \emph{compactification}
scheme to three spatial dimensions e.g. by using manifolds with non-trivial
\emph{fluxes} (unrelated to real magnetic fields). Different coupling of fermions
and bosons to such external fields breaks target space supersymmetry in a way independent of that induced by brane motion, which could be the dominant one in phenomenologically realistic models. The
consequent induced mass splitting~\cite{bachas,gravanis} between fermionic
and bosonic excitations on the brane world is proportional to the intensity
of the flux field (a string generalization of the well-known Zeeman effect of ordinary quantum mechanics, whereby the presence of an external field leads to energy splittings, which are however different  between (charged) fermions and bosons).
In this way one may obtain phenomenologically realistic mass
splittings in the excitation spectrum (at TeV or higher energy scales) owing
to \emph{supersymmetry obstruction} rather than spontaneous breaking (this terminology, which is due to E.~Witten~\cite{obstr}, means that,
although the ground state could still be characterised by zero vacuum energy, the masses of fermion and boson excitations differ and thus supersymmetry is broken at the level of the excitation-spectrum).
We also mention that the plausible assumption
of a population concentration of
massive D-particle defects in the haloes of galaxies can lead to modified galactic dynamics~%
\cite{recoil2}.

One can calculate the vacuum energy induced on the brane world
in such an adiabatic situation by considering its interaction with the D-particles as well as the
other branes in the construction. This calculation was presented in~\cite{emnw,emndvoid},
where we refer the interested reader for further details. Here we mention only the results
relevant for the present discussion. The important point to notice here is the fact that the forces exerted by the various structures, such as D-particles and other branes, on the brane world, representing the observable Universe in this kind of models, do not have a fixed sign.

We concentrate first on D0-particle/D8-brane interactions in the type-IA
model of~\cite{emnw}. During the late era of the
Universe when the approximation of adiabatic motion is valid, we use a weak-string-coupling
approximation $g_s \ll 1$. In such a case, the D-particle masses
$\sim M_s/g_s$ are large, \emph{i.e.}, these masses could be of the Planck size:
$M_s/g_s \sim M_P = 1.22 \cdot 10^{19}$~GeV or higher.
In the adiabatic approximation for the relative motion, these
interactions may be represented by a string stretched between the
D0-particle and the D8-brane, as shown in Fig.~\ref{dfoam:fig}. The world-sheet amplitude of
such a string yields the appropriate potential energy between the D-particle and the D-brane,
which in turn determines the relevant contribution to the vacuum energy of the brane.
As is well known~\cite{polch}, parallel relative motion does not generate any potential, and the only
non-trivial contributions to the brane vacuum energy come from motion transverse
to the D-brane. Neglecting a velocity-independent term in the D0-particle/D8-brane
potential that is cancelled for a D8-brane in the presence of orientifold $O_8$
planes~\cite{polch}~\footnote{This cancellation is crucial for obtaining
an appropriate supersymmetric string ground state with zero ground-state energy.}, we find~\cite{emnw}:
\begin{eqnarray}
\mathcal{V}^{short}_{D0-D8} & = & - \frac{\pi \alpha '}{12}\frac{v^2}{r^3}~~{\rm for} ~~
r \ll \sqrt{\alpha '}~,
\label{short}\\
\mathcal{V}^{long}_{D0-D8} & = & + \frac{r\, v^2}{8\,\pi \,\alpha '}~~~{\rm for} ~~
r \gg \sqrt{\alpha '}~.
\label{long}
\end{eqnarray}
where $v \ll 1$ is the relative velocity between the D-particle and the brane world, which is
assumed to be non-relativistic.
We note that the sign of the effective potential changes between short distances (\ref{short})
and long distances (\ref{long}). We also note that there is a minimum distance given by:
\begin{equation}
r_{\rm min} \simeq \sqrt{v \,\alpha '}~, \qquad v \ll 1~,
\label{newmin}
\end{equation}
which guarantees that (\ref{short}) is less than $r/\alpha'$, rendering
the effective low-energy field theory well-defined. Below this minimum distance,
the D0-particle/D8-brane string amplitude diverges
when expanded in powers of $(\alpha')^2 v^2 /r^4$. When they are separated from
a D-brane by a distance smaller than $r_{\rm min}$, D-particles should be considered as lying on
the D-brane world, and two D-branes separated by less than $r_{\rm min}$
should be considered as coincident.

We now consider a configuration with a moving D8-brane located at distances $R_{i}(t)$
from the orientifold end-planes, where $ R_1(t) + R_2(t) = R_0$ the fixed extent of the ninth bulk dimension, and the 9-density of the D-particles in the bulk is  denoted by $n^\star (r) $: see
Fig.~\ref{dfoam:fig}. The total D8-vacuum-energy density $\rho^8$ due to the relative motions
is~\cite{emnw}:
\begin{eqnarray}
&&  \mathcal{\rho}^{D8-D0}_{\rm total} = -  \int_{r_{\rm min}}^{\ell_s}  n^\star (r) \, \frac{\pi \alpha '}{12}\frac{v^2}{r^3}\, dr -  \int_{-r_{\rm min}}^{-\ell_s}  n^\star (r) \, \frac{\pi \alpha '}{12}\frac{v^2}{r^3}\, dr
   +  \nonumber \\ &&  \int_{-\ell_s} ^{-R_{1}(t)} \,n^\star (r) \frac{r\, v^2}{8\,\pi \,\alpha '}\, dr
+ \int_{\ell_s} ^{R_{2}(t)} \,n^\star (r) \frac{r\, v^2}{8\,\pi \,\alpha '}\, dr + \rho_0
\label{total}
\end{eqnarray}
where the origin of the $r$ coordinate is placed on the 8-brane world and $\rho_0$
combines the contributions to the vacuum energy density from inside the band
$-r_{\rm min} \le r \le r_{\rm min}$, which include the brane tension.
When the D8-brane is moving in a uniform bulk distribution of D-particles,
we may set $n^*\star(r) = n_0$, a constant, and the
dark energy density $\mathcal{\rho}^{D8-D0}_{\rm total}$ on the D8-brane is also
(approximately) constant for a long period of time:
\begin{equation}
\mathcal{\rho}^{D8-D0}_{\rm total} =  -n_0 \frac{\pi }{12} v(1 - v) + n_0 v^2 \frac{1}{16\pi \alpha '} (R_1(t)^2 + R_2(t)^2 - 2 \alpha ')
+ \rho_0~.
\label{total2}
\end{equation}
Because of the adiabatic motion of the D-brane, the time dependence of $R_i(t)$ is
weak, so that there is only a weak time dependence of the D-brane vacuum energy density:
it is positive if $\rho_0 > 0$, which can be arranged by considering branes with positive tension.

However, one can also consider the possibility that the D-particle density $n^\star (r)$
is inhomogeneous, perhaps because of some prior catastrophic cosmic collision,
or some subsequent disturbance. If there is a region depleted by D-particles~\cite{emndvoid} - a {\it D-void} -
the relative importance of the terms in (\ref{total2}) may be changed.
In such a case, the first term on the right-hand side of (\ref{total2}) may
become significantly  smaller than the term proportional to $R_i^2(t)/\alpha'$.
As an illustration, consider for simplicity and concreteness a situation in which there
are different densities of D-particles close to the D8-brane ($n_{\rm local}$) and at long distances to
the left and right ($n_{\rm left}, n_{\rm right}$). In this case, one obtains  from (\ref{total}):
 \begin{eqnarray}
&& \mathcal{\rho}^{D8-D0}_{\rm total} \simeq  -n_{\rm local} \frac{\pi }{12} v(1 - v) + n_{\rm left} v^2 \frac{1}{16 \pi \alpha '} \left(R_1(t)^2 - \alpha '\right)+  \nonumber \\ && n_{\rm right} v^2 \frac{1}{16 \pi \alpha '} \left( R_2(t)^2 - \alpha '\right)
+ \rho_0~.
\label{total3}
\end{eqnarray}
for the induced energy density on the D8-brane. The first term can be significantly smaller in
magnitude than the corresponding term in the uniform case, if the local density of D-particles
is suppressed. Overall, the D-particle-induced energy density on the D-brane world
{\it increases} as the brane enters a region where the D-particle density is {\it depleted}.
This could cause the onset of an \emph{accelerating phase} in the expansion of the Universe. It is
intriguing that the red-shift of GRB 090510~\cite{grb090510} is in the ballpark of the redshift range where the
expansion of the Universe apparently made a transition from deceleration to acceleration~\cite{decel}.
According to the above discussion, then, within our string foam model this \emph{may not} be a coincidence~\cite{emndvoid}.
The result (\ref{total3}) was derived in  an oversimplified case, where the possible
effects of other branes and orientifolds were not taken into account. However, as we argued in \cite{emndvoid},
the ideas emerging from this simple example persist in more realistic structures.

The important point for our purposes in such models of space-time foam, is the fact that there is an induced refractive index \emph{in vacuo}, as a consequence of  the photon interactions with the D-particles that cross the moving brane Universe. It is the \emph{linear density} of the D-particle defects $n(z)$ encountered by a propagating photon
that determines the amount of refraction. The density of D-particles crossing the D-brane
world cannot be determined from first principles,
and so may be regarded as a parameter in phenomenological models.
The flux of D-particles is proportional also to the velocity $v$ of the D8-brane in the bulk,
if the relative motion of the population of D-particles is ignored.

In order to make some phenomenological headway,
we adopt some simplifying assumptions. For example, we may assume that between
a redshift $z < 1$ and today ($z=0$), the energy density  has
remained approximately constant, as suggested by the available cosmological data.
As shown in \cite{emndvoid}, this assumption implies certain restrictions for the bulk density $n_{\rm short} (z)$ of D-particle defects near the brane world
\begin{equation}
\label{nofz}
n_{\rm short} (z) = n_{\rm short}(0) - \frac{12}{2^{7}\pi^{10}{\alpha '}^5} \frac{1}{\ell_s^9}\,\frac{c}{\ell_s}\, \frac{(v/c)^4}{(1 - \frac{v}{c})}\int_0^z \frac{d z'}{H(z') \, (1 + z')} \; \; {\rm where} \; \;  \ell_s \equiv \sqrt{\alpha '}~.
\end{equation}
We can make use of this result when we fit the available data from MAGIC and FERMI telescopes to this model.
For instance, by requiring that this density falls from O(1) at the redshifts $z=0.03$, relevant to MAGIC observations, by at least two orders of magnitude as we approach the red-shift $z=0.9$ of GRB 090510, we may constrain the relative motion of the D3-brane world. In fact, as discussed in \cite{emndvoid}, this can be achieved for relatively small velocities of the D3-brane world in the bulk, $v < 10^{-4} c$, consistent with the Early Universe Cosmology of the model~\cite{brany}.

Before closing this section  we would like to make a comment regarding the modification of (thermal) Dark Matter (DM) relic abundances in the foam model. As discussed in \cite{vergou}, the quantum fluctuations of the D-particles act as sources of particle production and thus affect the respective Boltzmann equation determining the relic abundance of DM particles. Additional modifications to this equation are due to the Finsler-like~\cite{finsler} character of the induced space-time metric during the interaction of neutral particles (like the DM ones) with the foam, due to its dependence on D-particle recoil velocities (and hence momentum transfer) (\emph{c.f}. (\ref{opsmetric2}) below).
We assume foam types in which the Lorentz symmetry is conserved on the average, and is violated only through fluctuations, namely we assume statistical ensembles of D-particles such that
$\ll u_i \gg = 0, \quad \ll u_i u_j \gg \equiv \frac{g^2_s}{M_s^2}\Delta_i^2 \overline{p_i}\, \overline{p}_j \delta_{ij}$, where $\overline{p}$ is some average momentum scale. One finds~\cite{vergou}:
\begin{eqnarray}\label{finalrelic}
\frac{\Omega_{\chi}'h_{0}^{2}}{(\Omega_{\chi}h_{0}^{2})_{\rm no~source}} \simeq
\left[ 1 +207.38g_s^2\frac{m^2}{M_s^2} x_f^{-2} \left(\sum_{i=1}^3 \Delta_i^2 \right)\right]^{1/2} \left[
1+ g_s^2 \frac{m^2}{M_s^2} \, \left(\sum_{i=1}^3 \Delta_i^2\right) \left(1+6x_0^{-1}\right)\right]
\end{eqnarray} where $x=m/T$, $T$ is the temperature, the suffix $f$ denotes the freeze-out point of the DM species, the suffix $0$ denotes present-day quantities, $\Omega_\chi' h_0^2$ denote the Hubble-constant-free present-day thermal relic abundance of a weakly interacting heavy Dark Matter particle of mass $m$, in the presence of the foam, and the suffix ``no source''
indicates the corresponding quantity in the Standard (foam-free) Cosmology.
Clearly the dominant correction terms are of order
$ g_s^2\frac{m^2}{M_s^2} \left(\sum_{i=1}^3 \Delta_i^2 \right) > 0$.

The modifications are suppressed by the square of the string scale (actually the D-particle mass $M_s/g_s$) and hence for relatively high string scales (much higher than TeV, of interest to us here) are small and do not lead to significant constraints on the density of foam. Indeed, in our case, from the MAGIC experiment, if we explain the observed delays as being due to the stringy foam alone, we obtain $M_s/g_s \sim 10^{19}$ GeV, while phenomenologically realistic DM candidates have masses $m$ in the range of at most a few TeV (usually a few hundreds of GeV)~\cite{lmn}.

It is important in this latter respect to make a comment regarding the nature of our foam: the D-particles in our approach~\cite{emnw,emnnewuncert,li} are viewed as \emph{background configurations} and \emph{not} as \emph{excitations} of the string vacuum. In other string/brane models, some authors have viewed the D-particles as localized excitations of the vacuum~\cite{dmatter}. In those cases, the D-particles  may be considered as dark matter candidates themselves and their density would be constrained by the cosmological observations on the DM sector to avoid overclosure of the Universe. As such they could not contribute to the refractive index.
Indeed, for the latter property to occur, light must interact coherently with the D-particles,
rather than scattering on them individually. Otherwise, there
would not just be a time delay  and thus an index of refraction, but the
light would be incoherently deflected at arbitrarily large angles.
Coherent scattering can only occur if the wavelength of light is
much greater than the mean separation of the scatterers. In
case the D0 particles behave like dark matter~\cite{dmatter}, with masses
near the Planck scale (to account for the MAGIC delays~\cite{MAGIC2}), their number density must be less than
$10^{-20}$ m$^{-3}$ to avoid overclosure of the Universe. Thus their mean
separation would be greater than $100$ m. The gamma rays that are of interest to us here
have a much shorter wavelength (smaller than cm) and therefore
cannot experience refraction due to these D-particles~\footnote{At any rate, such super-heavy DM would have been washed out by inflation in any realistic cosmology, so the scenario of \cite{dmatter} for D-particle excitations to play the r\^ole of DM pertains to much lighter D-particles in theories with low string mass scales.}.
In our D-foam model~\cite{emnw,emnnewuncert}, where the D0-branes (or the compactified D3 branes around 3-cycles in the model of \cite{li}) are viewed only as background defects, such an issue does not arise, and the density of the foam cannot be constrained by overclosure of the Universe issues. As we have discussed above, in this case the D-particles contribute to the dark energy sector (when in motion) consistently with current cosmological data. The cancellation between attractive (gravitational) and repulsive flux forces guarantees a supersymmetric vacuum, with zero vacuum energy,  if no relative motion of foam occurs.

In the above picture, the presence of a ground-state D-particle defect would be reflected in the world-sheet boundary conditions for the open strings representing radiation or matter excitations.
Thus, the only effects a recoiling D-particle would have can be summarized through its induced Finsler-like metric distortions
and the associated time delays of the interacting neutral matter or radiation, which we now proceed to discuss.

\section{Time Delays in D-particle foam and Stringy Uncertainties \label{sec:uncert}}
\begin{figure}[ht]
 \begin{center} \includegraphics[width=7cm]{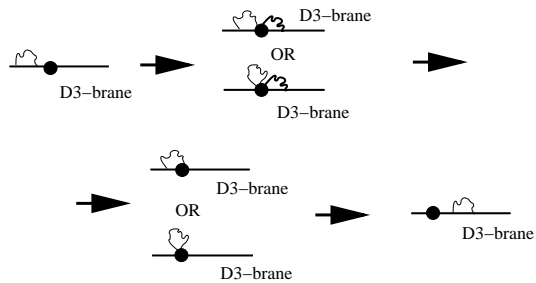}
\end{center}\caption{Schematic view of the capture process of an open string state, representing a photon propagating on a D3-brane world, by a D-particle on this world. The intermediate string state, indicated by thick wave lines, which is created on capture of the end(s) of the photon by the D-particle, stretches between the D-particle and the brane world, oscillates in size between $0$ and a maximal length of order $\alpha ' p^0$, where $p^0$ is the energy of the incident photon, and thus produces a series of outgoing photon waves, with attenuating amplitudes, constituting the re-emission process. The intermediate string state provides also the restoring force, necessary for keeping the D-particle roughly in its position after scattering.}
\label{fig:restoring}
\end{figure}

\vspace{0.1cm}
\paragraph{}
In this section we discuss a possible origin of time delays induced in the arrival time of photons, emitted simultaneously from an astrophysical object, as a result of their propagation
in the above-described D-particle space-time foam model. This comes by considering local interactions of photons, viewed as open string states in the model, with the D-particle defects.
As we shall argue, the above model is in principle capable of reproducing photon arrival time delays proportional to the photon energies, of the kind observed in the MAGIC experiment~\cite{MAGIC,MAGIC2}. This microscopic phenomenon, which is essentially stringy and does not \emph{characterize} local field theories, contributes to a sub-luminal non-trivial
refractive index \emph{in vacuo}, induced by the capture of photons or electrically neutral probes by the D-particle
foam~\cite{emnnewuncert}.
The capture process is described schematically in figure \ref{fig:restoring}.
An important feature of the model is that, on account of electric charge conservation, only electrically neutral excitations are subjected to capture by the D-particles. To the charged matter the D-particle foam looks \emph{transparent}. This is because the capture process of fig.~\ref{fig:restoring} entails a splitting of the open string matter state. Charged excitations are characterized by an electric flux flowing across the string, and when the latter is cut in two pieces as a result of its
capture by the D-particle defect, the flux should go somewhere because charge is conserved.
The D-particle, being neutral, cannot support this conservation, and hence only \emph{electrically neutral} excitations, such as photons, are subjected to this splitting and the associated delays. The reader should bear in mind of course that the D-particles carry other kinds of fluxes, unrelated to electromagnetism~\cite{polch}. These are conserved separately, and it is for this reason that isolated D-particles cannot exist, but there must always be in the company of other D-branes, as in our model above, so that the relevant fluxes are carried by the stretched strings between the latter and the D-particles.

For our purposes in this work it is also important to remark that the D-particles are treated as static when compared to photons. This is because the ends of the open string representing the photon move on the D3 brane world with the speed of light in (normal) vacuo, while the relative velocities of the D-particles with respect to the brane world (which propagates in the bulk space) are much lower than this. For instance, as discussed in \cite{brany}, to reproduce cosmological observations in this model, in particular the spectrum of primordial density fluctuations, which are affected by the relative motion of the brane world, the speed of propagation of the D3-brane Universe should be smaller than $10^{-4}c$.

When the end(s) of the open-string photon state are attached to the D-particle, there is an intermediate string state formed, \emph{stretched} between the D-particle and the D3-brane, which absorbs the incident energy $p^0$ of the photon state to grow in size from zero  to a maximum length $L$ that is determined by the requirement of energy minimization as follows: one assumes~\cite{sussk1} that the intermediate string state needs $N$ oscillations to achieve its maximal length $L$, as in the standard string-string scattering case. Following the same logic, we first observe that
energy conservation during the capture process of fig.~\ref{fig:restoring} leads to relations of the form
\begin{equation}
p^0 = \frac{L}{\alpha'} + \frac{N}{L}
\label{consenergy}
\end{equation}
Minimizing the right-hand-side with respect to $L$ and taking into account that the one end of the intermediate string state attached to the brane world moves with the \emph{speed of light in vacuo}, $c=1$ (in our units),  one
arrives at time delays for the outgoing photon waves of order~\cite{emnnewuncert}:
\begin{equation}
  \Delta t \sim  \alpha ' p^0~.
\label{delayonecapture}
\end{equation}
This delay is \emph{causal} and can also be obtained by considering scattering of open string states off D-particle backgrounds~\footnote{For type IA strings, which admit point-like D-particles, the situation is a straightforward extension of techniques
applied to the open string-string scattering case of \cite{sussk1}. For type IIB strings, the situation is technically
more involved, due to the compactified D3-branes, wrapped up around three cycles, that play the r\^ole of
``D-particles'' in the model~\cite{li}, but the results are qualitatively the same as in (\ref{delayonecapture}).}.
One assumes that the incident open string state splits into two open string states upon interaction with the D-particle defect, which, after their scattering in the D-particle background, the presence of which is implemented through the appropriate Dirichlet world-sheet boundary conditions of the open strings, re-join to produce the outgoing state.
The intermediate stage involves two open strings that scatter off each other.
The (quantum) oscillations of the intermediate string state will produce a series of outgoing wave-packets, with attenuating amplitudes, which will correspond to the re-emission process of the photon after capture.
The presence of the stretched string state, which carries the characteristic flux of the D-brane interactions, provides the \emph{restoring force}, necessary to keep the D-particle in its position after scattering with the photon.

The situation may be thought of as the stringy/brany analogue of the restoring force in situations in local field theories of photons propagating in media with non trivial refractive indices, as discussed by Feynman~\cite{feynman}. The r\^ole of the electrons in that case (represented as harmonic oscillators) is played here by the D-particle defects of the space-time. The stringy situation, however, is more complicated, since the D-particles have an infinite number of oscillatory excitations, represented by the various modes of open strings with their ends attached to them. Moreover, contrary to the conventional medium case, in the string model the refractive index is found proportional to the photon frequency, while the effective mass scale that suppresses the effect (\ref{delayonecapture}) is the quantum gravity (string) scale $M_s$. The latter property can be understood qualitatively by the fact that the mass of these D-particle defects is of order~\cite{polch} $1/(g_s~\sqrt{\alpha '}) = M_s/g_s$, where $g_s$ is the string coupling (in units of $\hbar = c = 1$).

We remark at this stage that the above time delays are a direct consequence of the \emph{stringy uncertainty principles}. Indeed, strings are characterized by two kinds of uncertainty relations:
the phase-space Heisenberg uncertainty, modified by higher order terms in $\alpha '$~\cite{venheisenberg} as a result of the existence of the minimal string length $\ell_s \equiv \sqrt{\alpha '}$ in target space-time,
\begin{equation}
  \Delta X \Delta P \ge \hbar  + \alpha ' (\Delta P)^2 + \dots
\label{heisenberg}
\end{equation}
and the time-space uncertainty relation~\cite{yoneya}
\begin{equation}
   \Delta X \Delta t \ge \alpha '~.
\label{stringyunc}
\end{equation}
Since the momentum uncertainty $\Delta P < p^0$, we have from (\ref{heisenberg}), to leading order in $\alpha '$ (in units $\hbar = 1$):
\begin{equation}
\Delta X \ge \frac{1}{\Delta P} > \frac{1}{p^0}~.
\end{equation}
In view of (\ref{delayonecapture}), we then arrive at consistency with the space-time uncertainty (\ref{stringyunc}),
\begin{equation}
\Delta X \ge \frac{\alpha'}{\alpha ' p^0} \sim \frac{\alpha '}{\Delta t}
\label{timespacenew}\end{equation}
As in the conventional string theory photon-photon scattering~\cite{sussk1}, these delays are \emph{causal}, \emph{i.e.} consistent with the fact that signals never arrive before they occur.
Hence they are \emph{additive} for multiple scatterings of photons by the foam defects from emission till observation.
 As we shall discuss in section~\ref{sec:totaldelay},  this provides
the necessary \emph{amplification}, so that the total delay of the more energetic photons can be~\cite{emnnewuncert} of the observed order in MAGIC and FERMI experiments.

In addition to this leading refractive index effect (\ref{delayonecapture}),
 there are corrections induced by the recoil of the D-particle itself, which contribute to space-time distortions that we now proceed to discuss.
From a world-sheet view point, the presence of $D$-particle recoil may be represented by adding to a fixed-point (conformal) $\sigma$-model action, the following deformation~\cite{recoil,szabo}:
\begin{equation}
\mathcal{V}_{\rm{D}}^{imp}=\frac{1}{2\pi\alpha '}
\sum_{i=1}^{D}\int_{\partial D}d\tau\,u_{i}%
X^{0}\Theta\left(  X^{0}\right)  \partial_{n}X^{i}. \label{fullrec}%
\end{equation}
where $D$ in the sum denotes the appropriate  number of spatial target-space dimensions.
For a recoiling D-particle confined on a D3 brane, $D=3$.

There is a specific type of conformal algebra, termed logarithmic conformal algebra~\cite{lcft}, that the recoil operators satisfy~\cite{recoil,szabo}.
This algebra is the limiting case of world-sheet algebras that can still be classified by conformal blocks. The impulse operator
$\Theta(X^0)$ is regularized so that the logarithmic conformal field theory algebra is respected~\footnote{This can be done by using the world-sheet scale, $\varepsilon^{-2} \equiv {\rm ln}\left(L/a\right)^2$, with $a$ an Ultra-Violet scale and $L$ the world-sheet area, as a regulator~\cite{recoil,szabo}: $\Theta_\varepsilon (X^0) = -i\,\int_{-\infty}^\infty \frac{d\omega}{\omega- i\varepsilon} e^{i\omega X^0}$. The quantity $\varepsilon \to 0^+$ at the end of the calculations.}.
The conformal algebra is consistent with momentum conservation during recoil~\cite{recoil,szabo}, which allows for the expression of the recoil velocity $u_i$ in terms of momentum transfer during the scattering
\begin{equation}
u_i = g_s\frac{p_1 - p_2}{M_s}~,
\label{recvel}
\end{equation}
with $\frac{M_s}{g_s}$ being the D-particle ``mass'' and $\Delta p \equiv p_1 - p_2$ the associated momentum transfer of a string state during its scattering  with the D-particle.

We next note that one can write the boundary recoil/capture operator (\ref{fullrec}) as a total derivative over the bulk of the world-sheet, by means of the two-dimensional version of Stokes theorem. Omitting from now on the explicit summation over repeated $i$-index, which is understood to be over the spatial indices of the D3-brane world, we write then:
\begin{eqnarray}\label{stokes}
&& \mathcal{V}_{\rm{D}}^{imp}=\frac{1}{2\pi\alpha '}
\int_{D}d^{2}z\,\epsilon_{\alpha\beta} \partial^\beta
\left(  \left[  u_{i}X^{0}\right]  \Theta_\varepsilon \left(  X^{0}\right)  \partial^{\alpha}X^{i}\right) = \nonumber \\
&& \frac{1}{4\pi\alpha '}\int_{D}d^{2}z\, (2u_{i})\,\epsilon_{\alpha\beta}
 \partial^{\beta
}X^{0} \Bigg[\Theta_\varepsilon \left(X^{0}\right) + X^0 \delta_\varepsilon \left(  X^{0}\right) \Bigg] \partial
^{\alpha}X^{i}
\end{eqnarray}
where $\delta_\varepsilon (X^0)$ is an $\varepsilon$-regularized $\delta$-function.
For relatively large times after the impact at $X^0=0$ (which we assume for our phenomenological purposes in this work), this is equivalent to a deformation describing an open string propagating in an antisymmetric  $B_{\mu\nu}$-background corresponding to an external constant in target-space ``electric'' field,
\begin{equation}
B_{0i}\sim u_i ~, \quad B_{ij}=0~, \quad (X^0 > 0)
\label{constelectric}
\end{equation}
where the $X^0\delta (X^0)$ terms in the argument of the electric field yield vanishing contributions in the large time limit, and hence are ignored from now on.

To discuss the space time effects of a recoiling D-particle on an open  string state propagating on a D3 brane world, we should consider a $\sigma$-model in the presence of the B-field (\ref{constelectric}), which leads to mixed-type boundary conditions for open strings on the boundary $\partial \mathcal{D}$ of world-sheet surfaces with the topology of a disc. Absence of a recoil-velocity $u_i$-field leads to the usual Neumann boundary conditions, while the limit where $g_{\mu\nu} \to 0$, with $u_i \ne 0$, leads to Dirichlet boundary conditions.

In analogy with the standard string case in background electric fields~\cite{seibergwitten,sussk1},
one obtains a \emph{non-commutative space-time} if recoil of the D-particle is taken into account. This can be seen
upon considering commutation relations among the coordinates of the first quantised $\sigma$-model in the background
(\ref{constelectric}). As in the standard case of a constant electric background field, in the presence of a recoiling D-particle, the pertinent non commutativity is between time and the spatial coordinate along the direction of the recoil velocity field (for large times $t$ after the impact at $t=0$):
\begin{equation}
[ X^1, t ] = i \theta^{10} ~, \qquad \theta^{01} (= - \theta^{10}) \equiv \theta =  \frac{1}{u_{\rm c}}\frac{\tilde u}{1 - \tilde{u}^2}
\label{stnc2}
\end{equation}
where, for simplicity and concreteness, we assume recoil along the spatial $X^1$ direction. Thus, the induced non commutativity is consistent with the breaking of the Lorentz symmetry of the ground state by the D-particle recoil. The quantity $\tilde{u}_i \equiv \frac{u_i}{u_{\rm c}}$ and  $u_{\rm c} = \frac{1}{2\pi \alpha '}$ is the Born-Infeld \emph{critical} field.
The space-time uncertainty relations (\ref{stnc2}) are consistent with the corresponding space-time string uncertainty principle (\ref{stringyunc}).

Of crucial interest in our case is the form of the induced open-string \emph{effective target-space-time metric}.
The situation parallels that of open strings in external electric field backgrounds, discussed  in refs.~\cite{sussk1,seibergwitten}. Hence,
the effective open-string metric, $g_{\mu\nu}^{\rm open,electric}$, which is due to the presence of the recoil-velocity field $\vec{u}$, whose direction breaks target-space Lorentz invariance, is obtained by
extending appropriately the results of strings in constant electric field backgrounds~\cite{seibergwitten,sussk1} to the background (\ref{constelectric}):
\begin{eqnarray}
           g_{\mu\nu}^{\rm open,electric} &=& \left(1 - {\tilde u}_i^2\right)\eta_{\mu\nu}~, \qquad \mu,\nu = 0,1 \nonumber \\
           g_{\mu\nu}^{\rm open,electric} &=& \eta_{\mu\nu}~, \mu,\nu ={\rm all~other~values}~.
\label{opsmetric2}
\end{eqnarray}
For concreteness and simplicity, we considered a frame of reference where the matter particle
has momentum only across the spatial direction $X^1$, \emph{i.e.} $0 \ne p_1 \equiv p \parallel u_1~, p_2=p_3 =0$.
Moreover, as in the standard case of strings in an electric field background, there is a modified effective string coupling~\cite{seibergwitten,sussk1}:
\begin{equation}
   g_s^{\rm eff} = g_s \left(1 - \tilde{u}^2\right)^{1/2}
\label{effstringcoupl2}
\end{equation}
The fact that the metric in our recoil case depends on momentum transfer variables implies that D-particle recoil induces Finsler-type metrics~\cite{finsler}, \emph{i.e}. metric functions that depend on phase-space coordinates of the matter (photon) state.

We now mention that the presence of the critical ``background field'' $u_c$, at which both the metric (\ref{opsmetric2}) and the effective coupling  (\ref{effstringcoupl2}) vanish, is associated with the \emph{destabilization of the vacuum}~\cite{burgess} when the field intensity approaches the \emph{critical value}. Since in our D-particle foam case, the r\^ole of the `electric' field is played by the recoil velocity of the
D-particle defect, the critical field corresponds to the relativistic speed of light, in accordance with special relativistic kinematics, which is respected in string theory by construction.
On account of (\ref{recvel}), then, this implies an upper bound on the induced momentum transfer, and hence on the available momenta, for the effective field theory limit to be valid. Indeed, if we represent $\Delta p$ in (\ref{recvel}) as a fraction of the incident momentum $p_1$,  $\Delta p = r p_1$, $r < 1$, then the condition that the recoil velocity of the D-particle is below the speed of light in vacuo, as required by the underlying consistency of strings with the relativity principle, implies~\cite{emncomment}
\begin{equation}\label{novelgzk}
\Delta p \equiv r p_1 < \frac{M_s}{g_s} \Rightarrow p_1 < \frac{M_s}{r\, g_s}~.
\end{equation}
When the incident momentum approaches the order of this cutoff, the effective string coupling (\ref{effstringcoupl2}) vanishes, while above that value the coupling becomes imaginary, indicating complete absorption of the string state by the D-particle. The space-time distortion due to recoil is so strong in such a case that there is no possibility of re-emergence of the string state, the defect behaves like a black hole, capturing permanently the string state.

It is important to notice that in modern string theory the quantities $M_s, g_s$ are completely phenomenological.
In fact, it is possible to construct phenomenologically realistic string/brane-Universe models, in the sense of being capable to incorporate the Standard Model particles at low energies, with string scales in such a way that
$M_s/g_s$ is significantly lower than the Planck scale. For instance, there are constructions~\cite{pioline} in the large extra dimension framework, for which
$M_s/g_s$ is of the order of $10^{19}$ eV, \emph{i.e}. of order of the conventional GZK cutoff in Lorentz invariant particle physics models. Thus, by embedding the D-particle foam to such models, one may have the appearance of a \emph{novel type} Gretisen Zatsepin Kusmin (GZK) cutoff~\cite{GZK}, of similar order to the one obtained from conventional Lorentz invariance arguments, but of quite different origin: here it is the \emph{subluminal} nature of the \emph{recoil velocity} of the foam that sets the new upper bound in momentum transfer. This is not fine-tuning in our opinion, but indicates the appearance of a new upper bound in momenta, related implicity to the underlying Lorentz invariance of the string theory, which is broken spontaneously by the recoiling D-particle background. This point will become of importance later on, when we discuss
constraints of our model coming from ultra-high-energy/infrared photon/photon scattering~\cite{sigl}.

Before closing this section we make an important remark. The induced metric (\ref{opsmetric2}) will affect the dispersion relations of the photon state:
\begin{equation}\label{drps}
p^\mu p^\nu g_{\mu\nu}^{\rm open,electric} = 0~.
\end{equation}
However, because the corrections on the recoil velocity $u_i$ are quadratic, such modifications will be suppressed by the square of the D-particle mass scale.
On the other hand, the presence of the D-particle recoil velocity will affect the induced time delays (\ref{delayonecapture})
 by higher-order corrections of the form, as follows by direct analogy of our case with that of open strings in  a constant electric field~\cite{sussk1}:
 \begin{equation}\label{timerecoil}
 \Delta t_{\rm with~D-foam~recoil~velocity} = \alpha ' \,\frac{p^0}{1 - {\tilde u}_i^2}~.
\end{equation}
Thus, the D-particle recoil effects are quadratically suppressed by the D-particle mass scales, since (\emph{c.f.} (\ref{recvel})) ${\tilde u}_i \propto g_s \Delta p_i /M_s$, with $\Delta p_i$ the relevant string-state momentum transfer.
However, the leading order delay effect (\ref{timerecoil}), obtained formally by considering the limit of vanishing recoil velocity, is linearly suppressed by the string scale, and thus the induced time delays are disentangled in the string foam model from the modified Finsler dispersion relations (\ref{drps}).
This is important for the phenomenology of the model, as we shall discuss below.
\vspace{0.1cm}

\noindent\emph{\textbf{Lack of a Local Effective Field Theory Formalism in recoiling D-particle Foam:}} An important comment arises at this stage concerning the construction of a possible local effective field theory action in the case of D-particle foam, when recoil of the D-particle defect is considered.
In view of the formal analogy of the problem with that of an external electric background field, one is tempted to apply the same considerations as those leading to the non-commutative (NC) effective actions in field theory, for instance
for case of NC quantum electrodynamics of ref.~\cite{carroll}. Naively, one might think of writing down a
Standard-Model-Extension type Lagrangian~\cite{kostelecky}, including non-commutativity (and thus Lorentz-violating but CPT conserving) terms in the presence of D-foam background, with the $\theta^{\alpha\beta}$ parameter being replaced by $\theta^{0i} = u_i = g_s \frac{\Delta p_i}{M_s}$, the D-particle recoil velocity.

However, this is not correct. As a result of the momentum-transfer dependence of the non-commutativity parameter in the D-foam case, it is not possible to write down the effective action in target space as a power series of local quantum operators. The momentum transfer is not represented by such operators when taking the Fourier transform.

Thus writing down a local effective action for the recoiling D-particle in interaction  with, say, a photon, is not possible. This has important consequences for phenomenology. Indeed, the time delays (\ref{timerecoil}), which are associated with the stringy uncertainty, are found proportional to the incident energy of the (split) photon state
and are thus linearly suppressed by the string scale. The associated refractive index is therefore linearly suppressed, but this cannot be interpreted as an average propagation of photons in the context of some local effective action. As we have seen above, the associated anomalous photon dispersion terms in this case, induced by the Finsler metric (\ref{opsmetric2}), are  \emph{quadratically} suppressed by the string scale.

Therefore, any analysis, such as those involving high energy cosmic rays~\cite{sigl} using an effective linear dispersion relation, obtained from a local effective action, does not apply to our problem. Moreover, it is a generic feature of any local effective theory, that is a theory on flat space times involving higher derivatives operators, to yield birefringence, since the corresponding modified dispersion relations for photons reduce at the end of the day to a solution of the photon frequency as a function of the wave number $k$ which is obtained from a quadratic equation. The latter admits two solutions with different propagation for the two photon polarizations.
This is \emph{not} the case of the uncertainty-related time delay (\ref{timerecoil}), which, as we have seen, is independent of the photon polarization~\cite{emnnewuncert}.

\section{Multiple Photon-D-particle Scatterings in the Foam and Total (observed) Time Delays \label{sec:totaldelay}}
\vspace{0.1cm}
\paragraph{}
The above-discussed time delays (\ref{delayonecapture}) pertain to a single encounter of a photon with a D-particle. In case of a foam, with a linear density of defects $n^*/\sqrt{\alpha '}$, \emph{i.e.} $n^*$ defects per string length, the overall delay encountered in the propagation of the photon from the source to observation, corresponding to a traversed distance $D$, is:
\begin{equation}
\Delta t_{\rm total} = \alpha ' p^0 n^* \frac{D}{\sqrt{\alpha '}} = \frac{p^0}{M_s} n^* D~,
\label{totaldelay}
\end{equation}
where $p^0$ denotes an average photon energy.
When the Universe's expansion is taken into account, one has to consider the appropriate red-shift-$z$ dependent stretching factors, which affect the measured delay in the propagation of two photons with different energies, as well as the Hubble expansion rate $H(z)$.
Specifically, in a Roberston-Walker cosmology the delay due to any single scattering
event is affected by: (i) a time dilation factor~\cite{JP} $(1 + z)$ and (ii) the redshifting~\cite{robust}
of the photon energy which implies that the observed energy of a photon
with initial energy $E$ is reduced to $E_{\rm obs} = E_0/(1 + z)$. Thus, the observed
delay associated with (\ref{delayonecapture}) is~\cite{robust,JP}:
\begin{equation}\label{obsdelay}
\delta t_{\rm obs} = (1 + z) \delta t_0 = (1 + z)^2 \sqrt{\alpha '} E_{\rm obs} .
\end{equation}
For a line density of D-particles $n(z)$ at redshift $z$, we have $n(z) d\ell =  n(z) dt$ defects
per co-moving length, where $dt$ denotes the infinitesimal Robertson-Walker time interval
of a co-moving observer. Hence, the total delay of an energetic photon in a co-moving
time interval $dt$ is given by $n(z) (1 + z)^2 C \sqrt{\alpha '} E_{\rm obs} \, dt $. The
time interval $dt$ is related to the Hubble rate $H(z)$ in the standard way:
$dt = - dz/[(1 + z) H(z)]$.  Thus, from (\ref{obsdelay}) we obtain a total delay in the arrival times of photons with energy difference $\Delta E$,  which has the form considered in \cite{mitsou,robust}, namely it is proportional to $\Delta E$ and is suppressed linearly by the quantum gravity (string) scale, $M_s$:
\begin{equation}
(\Delta t)_{\rm obs} =  \frac{ \Delta E}{M_s} {\rm H}_0^{-1}\int_0^z n^*(z) \frac{(1 + z)dz}{\sqrt{\Omega_\Lambda + \Omega_m (1 + z)^3}}
\label{redshift2}
\end{equation}
where $z$ is the red-shift, ${\rm H}_0$ is the (current-era) Hubble expansion rate, and we have
assumed for concreteness the $\Lambda$CDM standard model of cosmology. The reader should compare this relation
with (\ref{redshift}), based on naive considerations. The presence in (\ref{redshift2}) of the linear density of defects $n^*(z)$, which is in general red-shift dependent, is a crucial difference, and it is this feature that allows the string model to fit simultaneously the MAGIC result and the data from the short burst GRB 090510, as we shall discuss below.

Notice in (\ref{redshift2}) that the essentially stringy nature of the delay implies that the characteristic suppression scale is the string scale $M_s$, which plays the r\^ole of the quantum gravity scale in this case. The scale $M_s$ is a free parameter in the modern version of string theory, and thus it can be constrained by experiment. As we have discussed in this article, the observations of delays of energetic (TeV) photons from AGN by the MAGIC telescope~\cite{MAGIC} can provide such an experimental way of constraining $n^*/M_s$ in (\ref{redshift}). For $\Delta E \sim 10$ TeV, for instance, the delay (\ref{redshift}) can lead to the observed one of order of minutes, provided $M_s/n^* \sim 10^{18}$ GeV (in natural units with $c=1$)~\cite{MAGIC2}. This implies natural values for both $n^*$ and $M_s$, although it must be noted that $n^*$ is another free parameter of the bulk string cosmology model of \cite{emnw}, considered here. In general, $n^*(z)$ is affected by the expansion of the Universe, as it is diluted by it, but also depends on the bulk model and the interactions among the D-particles themselves. For inhomogeneous foam situations, the dependence of $n^*$ on the redshift can be quite complicated~\cite{emndvoid}. For redshifts of relevance to the MAGIC experiments, $z=0.034 \ll 1$, one may ignore the $z$-dependence of $n^*$ to a good approximation.

The total delay (\ref{redshift2}) may be thought of as implying~\cite{feynman} an effective \emph{subluminal} refractive index $n(E)$ of light propagating in this space time, since one may assume that the delay is equivalent to light being slowed down due to the medium effects.
 On account of the theoretical uncertainties in the source mechanism, however, the result of the
 AGN Mkn 501 observations of the MAGIC Telescope translate to \emph{upper} bounds for the quantity $n^*/M_s$ in (\ref{redshift}), which determines the strength of the anomalous photon dispersion in the string/D-particle foam model.

In view of the above discussion, if the time delays observed by MAGIC can finally be attributed partly or wholly to this type of stringy space-time foam, and therefore to the stringy uncertainties, then the AGN Mkn 501, and other such celestial sources of very high energy photons, may be viewed as playing the r\^ole of Heisenberg microscopes and amplifiers for the stringy space-time foam effects.

But how, then, can we incorporate the data from GRB 0905210~\cite{grb090510} in this model?
It is clear that a fit with a linearly modified dispersion relation  does not work in this case for the values of the Quantum Gravity scale $M^{\rm MAGIC}_{\rm QG\,1} \sim 10^{18}$~GeV that fit the MAGIC, H.E.S.S. and the other FERMI data. Indeed, the observed short delays of this burst can be explained on the basis of linearly modified dispersion relations only if the quantum gravity scale is \emph{larger} than $M_{\rm QG\,1} \simeq 1.2 M_P$.

As we have already mentioned, this argument has been used in \cite{grb090510} in order to exclude all models of QG involving linearly suppressed time delays for photons, on account of naturalness.
We would not agree with this statement. Leaving aside the fact that from a single measurement, with uncertainties on the precursor of the GRB, one cannot draw safe conclusions, we mention that what one calls a \emph{natural
scale of QG  } is highly model dependent. For instance, as we have seen above,
in the string foam model,  the relevant ``QG scale'', that dictates the order of the foam-induced time delays of photons, is a complex function of the model's parameters and is not simply given by the string scale $M_s$~\cite{emndvoid}. Indeed, as becomes clear from (\ref{totaldelay}) (or (\ref{redshift})) the relevant scale is not simply the D-particle mass, $M_s/g_s$, but a combination
\begin{equation}\label{effqgscale2}
M_{\rm QG-D-foam} \sim \frac{M_s}{g_s\, n^*(z)}
\end{equation}
 involving the linear number density $n^*(z)$ of the foam defect, encountered by the photon during its propagation from the source till observation. This quantity depends on the bulk density of the D-particles, which in the model of \cite{emnw} is a free parameter.

Inhomogeneous bulks are perfectly consistent background configurations for our brane world scenario~\cite{emndvoid}.
In order to match the photon delay data of Fig.~\ref{fig:data} with the D-foam model,
we need a reduction of the \emph{linear} density of defects encountered by the photon by
about two orders of magnitude in the region $0.2 < z < 1$,
whereas for $z < 0.2 $ there must be, on average, one D-particle defect per unit string
length $\ell_s$. We therefore assume that our D-brane encountered a D-void when $0.2 < z < 1$,
in which there was a significant reduction in the bulk nine-dimensional density of defects.
Such assumptions can be consistent with cosmological considerations on the dark sector of the model, as explained in \cite{emndvoid}. In particular, an important parameter for the cosmology of the model is the propagation velocity $v$ of the brane world in the bulk (\emph{c.f}. fig.~\ref{dfoam:fig}). As we have discussed previously (\ref{nofz}) this is related~\cite{emndvoid} to the density of the defects near the brane world, $n^{\rm short}(z)$, and hence to the linear density of defects encountered by the photon on the brane world.

To have a reduced density by two order of magnitude at redshifts $z=0.9$, while having a density of defects of $O(1)$ per string length at redshifts $z < 0.1$, one  must consider
the magnitude of $v$ as well as the string scale $\ell_s$.
For instance, it follows from (\ref{nofz}) that for string energy scales of the order of TeV, \emph{i.e.} string time scales $\ell_s/c = 10^{-27}~s$, one must consider a brane velocity $v \le \sqrt{10} \times 10^{-11} \, c$,
which is not implausible for a slowly moving D-brane at a late era of the Universe~\footnote{Much
smaller velocities are required for small string scales that are
comparable to the four-dimensional Planck length.}.
This is compatible with the constraint on $v$ obtained from inflation in~\cite{brany},
namely $v^2 \le 1.48 \times 10^{-5}\,g_s^{-1}$, where $g_s < 1$ for the
weak string coupling we assume here.
On the other hand, if we assume a 9-volume $V_9 = (K \ell_s)^9$:
$K \sim  10^3$ and $\ell_s \sim  10^{-17}/{\rm GeV}$, then the brane velocity $v \le 10^{-4} c$.
In  our model, due to the friction induced on the
D-brane by the bulk D-particles, one would expect that the late-epoch brane velocity
should be much smaller than that during the inflationary era immediately following
a D-brane collision~\cite{brany}.

We mention for completion, that the above considerations pertain strictly speaking to
point-like D-particles, which are allowed only in type IA string constructions. However, as discussed in \cite{emndvoid}, similar conclusions can be applied to type-IIB constructions of D-foam~\cite{li}.

\section{Other (astrophysical) constraints on quantum-gravity foam \label{sec:biref}}
\vspace{0.1cm}
\paragraph{}
The sensitivity of the MAGIC (and FERMI) observations to Planck scale physics, at least
for linearly suppressed modified dispersion relations, calls for an immediate comparison with other sensitive probes of non-trivial optical properties of QG medium.

Indeed, from the analysis of \cite{MAGIC2}, there was no microscopic model dependence of
the induced modifications of the photon dispersion relations, other than the sub-luminal nature of the induced refractive index and the associated absence of birefringence, that is the independence
of the refractive index on the photon polarization. The latter feature avoids the otherwise very stringent constraints on the photon dispersion relation imposed by astrophysical observations, as we now come to discuss.

We shall be very brief in our description of the complementary astrophysical tests on Lorentz invariance and quantum-gravity modified dispersion relations, to avoid large diversion from our main point of this review article which is string theory.

There are three major classes of complementary astrophysical constraints, to be considered in any attempt to interpret the MAGIC, FERMI or more general $\gamma$-ray Astrophysics results in terms of quantum-gravity induced anomalies in photon dispersion.

\begin{itemize}

\item{\emph{\textbf{Birefringence and strong constraints on QG-induced photon dispersion}}}

In certain models of quantum gravity, with modified dispersion relations, for instance the so-called loop-quantum gravity~\cite{gambini}, the ground state breaks reflexion symmetry (parity) and this is one of the pre-requisites for a dependence of the induced refractive index on the photon polarization, \emph{i.e}. birefringence. We remind the reader that in birefringent materials this is caused precisely by the existence of some kind of anisotropies in the material. The velocities of the two photon polarizations (denoted by $\pm$) in such QG models may be parametrised by:
\begin{equation}
v_{\pm} = c\left(1 \pm \xi (\hbar \omega /M_P )^n \right)
\label{birefrrel}
\end{equation}
where $M_P = 1.22 \times 10^{19}$ GeV is the Planck energy scale, and $\xi$ is a parameter
following from the underlying theoretical model, which is related with the modifications
of the pertinent dispersion relations for photons. The order of suppression
of these effects is described by $n$ which in the models of \cite{gambini} assumed the value $n=1$, but in general one could have higher order suppression.

 Vacuum QG birefringence should have showed up in optical measurements from remote astrophysical sources, in particular GRBs.
Ultraviolet (UV) radiation measurements from distant galaxies~\cite{uv} and UV/optical polarization measurements of light from  Gamma Ray Bursters~\cite{grb} rule out
birefringence unless it is induced at a scale (way) beyond the Planck mass (for linear models,
the lower bound on the QG scale in such models can exceed the Planck scale ($\sim 10^{19}$ GeV) by as much as \emph{seven orders of magnitude}). Indeed,  in terms of the parameter $\eta$ introduced above (\emph{c.f.} (\ref{birefrrel})), for the case $n=1$ of \cite{gambini}, one finds from
 optical polarization observations that the absence of detectable birefringence effects imply the upper bound  $|\xi| < 2 \times 10^{-7}$, which is incompatible with the MAGIC observed delays,
 saturating from below the bounds imposed by the MAGIC experiment.

 At this point, we wish to mention that, using recent polarimetric observations of the Crab Nebula in the hard X-ray band by INTEGRAL~\cite{integral}, the authors of \cite{macio} have demonstrated  that the absence of vacuum birefringence effects constrains linearly suppressed Lorentz violation in quantum electrodynamics to the level $|\xi | < 6 \times 10^{-10}$ at 95\% C.L., thereby tightening by about three orders of magnitude the above-mentioned constraint.

\item{\emph{\textbf{Synchrotron Radiation and further stringent constraints for electronic QG-induced anomalous dispersion in vacuo}}}

Another important experimental constraint on models with QG-induced anomalous dispersion relations
comes from observations of synchrotron radiation from distant galaxies~\cite{crab,ems,crab2}, such as Crab Nebula (\emph{c.f.} fig.~\ref{fig:crab}). As well known, the magnetic fields at the core regions of galaxies curve the paths of (and thus accelerate) charged particles, in particularly electrons (which are stable and therefore appropriate for astrophysical observations), and thus, on account of energy conservation this results in synchrotron radiation.

\begin{figure}[H]
\begin{center}
\includegraphics[width=0.4\linewidth]{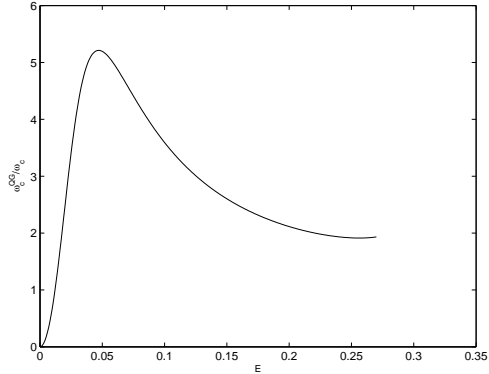}
\end{center}
\caption{The Crab Nebula is a supernova remnant, with a rotating neutron star (the Crab Pulsar) at its centre (middle image). Observations of synchrotron radiation from such celestial objects place very stringent constraints on quantum gravity models with anomalous dispersion relations for electrons (from the first of Ref.~\protect\cite{ems}).}
\label{fig:crab}
\end{figure}

In standard electrodynamics~\cite{jac}, electrons in an
external magnetic field ${H}$, follow helical orbits
transverse to the direction of ${H}$. The so-accelerated electrons in a magnetic field
emit synchrotron radiation with a spectrum that cuts off sharply at a
frequency $\omega_c$ (\emph{c.f.} fig.~\ref{fig:crab}, right panel):
\begin{equation}
\omega^{LI}_c = \frac{3}{2}\frac{eH}{m_0}\frac{1}{1 - \beta^2},
\label{sync}
\end{equation}
where $e$ is the electron charge, $m_0$ its mass, and $\beta_\perp \equiv v_\perp $ is the component of the velocity of the electron perpendicular to the direction of the
magnetic field.
The superfix $LI$ in (\ref{sync})
stresses that this formula is based on a LI approach, in which one
calculates the electron trajectory in a given magnetic field $H$ and the
radiation produced by a given current, using the relativistic relation
between energy and velocity.

All these assumptions are
affected by violations of Lorentz symmetry, such as those encountered in quantum-gravity space-time foam models, leading to modified dispersion relations
of the form:
\begin{eqnarray}
 \omega^2(k)&=& k^2 + \xi_\gamma \frac{k^{2 + \alpha}}{M_P^\alpha},
 \label{eq:pdr}\\
 E^2(p)&=& m_0^2+ p^2 + \xi_e \frac{p^{2 + \alpha}}{M_P^\alpha},
 \label{eq:mdr}
\end{eqnarray}
for photons (\ref{eq:pdr}) and electrons (\ref{eq:mdr}), where $\omega$ and $k$ are the
photon frequency and wave number, and $E$ and $p$ are the electron energy
and momentum, with $m_0$ the electron (rest) mass. In the spirit of the MAGIC observation analysis above, we assume here linear ($\alpha =1$) or quadratic QG ($\alpha = 2$)
effects, characterized by parameters $\xi_\gamma$ and $\xi_e$, extracting
the Planck mass scale $M_{P}=1.22\times 10^{19}$ GeV. In fact one can do the analysis~\cite{ems} for a general $\alpha$ (single power) and attempt to extract limits on this parameter by matching with observations.

A detailed analysis~\cite{crab,ems}, including the modifications in the electron's trajectories due to space-time foam~\cite{ems}, yields:
\begin{eqnarray}
&& \omega_c^{QG} \propto \omega_c^{LI} \frac{1}{
(1 + \sqrt{2 -1/{\cal \eta}^2})^{1/2}
\left(\frac{m_0^2}{E^2} + (\alpha + 1)\left(\frac{E}{{\cal M}}\right)^\alpha
\right)}~, \nonumber \\
&& {\cal M} \equiv M_P/|\xi_e|~, \quad \eta \equiv
1 - (E/{\cal M})^\alpha~,
\label{modqg}
\end{eqnarray}
where $\omega_c^{LI}$ is given in (\ref{sync}) and the superscript ``QG'' indicates that the QG-modified dispersion relations (\ref{eq:mdr})
are used. This function is plotted schematically (for $\alpha =1$)
in fig.~\ref{fig:crab} (right panel).

In \cite{crab,ems},
the above QG-modified dispersion relations have been tested using observations
from Crab Nebula.
It should be emphasized that the
estimate of the end-point energy of the Crab synchrotron spectrum
and of the magnetic field used above are indirect values based on
the predictions of the Synchrotron Self-Compton (SSC) model
of very-high-energy emission from Crab Nebula~\cite{reports}. In \cite{ems}
the choice of parameters used was the one that gives good agreement
between the experimental data on high-energy emission and the
predictions of the SSC model \cite{reports,hillas}.
Estimating the magnetic
field of Crab Nebula in the region
$160 \times 10^{-6}~{\rm Gauss} ~ < ~ H ~ < ~ 260 \times 10^{-6} ~{\rm Gauss}$,
and requiring $|\xi_e| \le 1$ (which thus sets the quantum gravity scale as at least $M_P$)
one obtains
the following bounds on the exponent $\alpha $ of the dispersion relations (\ref{eq:mdr})~\cite{ems}:
\begin{equation}
\alpha \ge \alpha_c~: \qquad 1.72 < \alpha_c < 1.74
\label{alphabounds}
\end{equation}
These results
imply already a sensitivity
to quadratic QG corrections with Planck mass suppression $M_P$.

However, for photons there are no strong constraints on $\xi_\gamma$ coming from synchrotron radiation studies, unless in cases where QG models entail  birefringence~\cite{crab2}, where, as we discussed above, strong constraints on the photon dispersion are expected at any rate from optical measurements on GRBs.
In this sense, the result (\ref{alphabounds}) \emph{excludes the possibility} that the MAGIC observations
leading to a four-minute delay of the most energetic photons are due to a quantum foam that acts \emph{universally} among photons and electrons. However, the synchrotron radiation measurements cannot exclude anomalous photon dispersion with linear Planck-mass suppression, leading to a saturation of the lower bound imposed by the MAGIC experiment~\cite{MAGIC2}, in models where the foam is transparent to electrons, as in the string foam case.

\item{\emph{\textbf{Strong constraints from Ultra-high-energy Cosmic photon annihilation}}}

Further strong constraints on generic modified dispersion relations for photons, like the ones used in the aforementioned QG-interpretation of the MAGIC results~\cite{MAGIC2}, comes from processes of scattering of ultra-high-energy photons, with energies above $10^{19}$ eV, off
very-low energy cosmic photons, such as the ones of the cosmic microwave background (CMB) radiation that populates the Universe today, as a remnant from the Big-Bang epoch.
In \cite{sigl} it has been argued that the non-observation of such ultra-high energy (UHE) photons
places very strong constraints on the parameters governing the modification of the photon dispersion relations, that are several orders of magnitude smaller than the values required to reproduce the MAGIC time delays, should the effect be attributed predominantly to photon propagation in a QG dispersive medium.

The main argument relies on the fact that
an ultra-high-energy photon would interact with a low-energy (``infrared'') photon of the CMB background (with energies in the eV range) to produce electron prositron pairs, according to the reaction:
\begin{equation}
  \gamma_{\rm UHE} \quad + \quad \gamma_{\rm CMB} \quad \Rightarrow \quad e^+~e^- ~.
\label{heirreact}
\end{equation}
The basic assumption in the analysis is the strict energy and momentum conservation in the
above reaction, despite the modified dispersion relations for the photons.
Such an assumption stems from the validity of a \emph{local-effective-lagrangian description}
of QG foam effects on particles with energies much lower than the QG energy scale (assumed close to Planck scale $M_{\rm Pl} = 10^{19}$ GeV). In this formalism, one can represent effectively the foam \emph{dispersive effects}  by higher-derivative \emph{local operators} in a \emph{flat-space-time} Lagrangian. The upshot of this is the modification of the pertinent equations of motion for the photon field (which in a Lorentz-invariant theory would be the ordinary Maxwell equations) by higher-derivative terms, suppressed by some power of the QG mass scale.

One considers the modified dispersion relations (\ref{eq:pdr}), (\ref{eq:mdr}), which in the notation of \cite{sigl}, taking explicit account of the various polarizations and helicities, can be written as:
\begin{eqnarray}\label{siglmdr}
  && \omega^2_{\pm} = k^2 + \xi_n^{\pm} k^2 \left(\frac{k}{M_{P}}\right)^n~, \qquad \omega_b^2 = k_b^2~, \nonumber \\
 &&  E_{e,\pm}^2 = p_e^2 + m_e^2 + \eta^{e,\pm}_n p_e^2 \left(\frac{p_e}{M_{P}}\right)^n~
\end{eqnarray}
with $(\omega, \vec k)$ the four-momenta for photons, and $(E, \vec{p}_e )$ the corresponding four-momentum vectors for electrons; the suffix $b$ indicates a low-energy CMB photon, whose dispersion relations are assumed approximately the normal ones, as any QG correction is negligible due to the low values of energy and momenta. The +(-) signs indicate left(right) polarizations (photons) or helicities (electrons). Positive (negative) $\xi$ indicate subluminal (superluminal) refractive indices.
Upon the assumption of energy-momentum conservation in the process,  one arrives at kinematic equations for the threshold of the reaction (\ref{heirreact}), that is the minimum energy of the high-energy photon required
to produce the electron-positron pairs.

For the linear- or quadratic- suppression case,  for which $n=1,2$ respectively (\ref{siglmdr}), one finds that, for the relevant subluminal photon refractive indices corresponding to the saturation of the lower bound on the QG scale $M_{QG1} \sim 10^{18}~{\rm GeV}$, the threshold for pair production disappears for ultra-high-energy photons, and hence such photons should have been observed.
The non-observation of such photons implies constraints for the relevant parameters $\xi, \eta$
which are stronger by \emph{several orders of magnitude} than the bounds on the QG scale inferred from the MAGIC observations.

From the analysis of \cite{sigl} one concludes that in the case of linear Planck-mass suppression of the sub-luminal QG-induced modified dispersion relations for photons, of interest for   the QG-foam interpretation of the MAGIC results~\cite{MAGIC2}, parameters with size $\xi_1 > 10^{14}$ are ruled out. This exceeds the sensitivity of the MAGIC experiment to such Lorentz-symmetry violating effects by fifteen orders of magnitude !
Similar strong constraints are also obtained from the non observations of \emph{photon decay} ($\gamma \to e^+ e^-$), a process which, if there are modified dispersion relations, is in general allowed~\cite{sigl}.

\end{itemize}

\section{The String Foam Models Evade the above Constraints}
\vspace{0.1cm}
\paragraph{}
From the above discussion it becomes clear that any model
of refraction in space-time foam that exhibits effects at the level of the MAGIC experiment sensitivity~\cite{MAGIC2} should be characterised by the following specific properties:
\begin{itemize}

\item{(i)} photons are \emph{stable} (\emph{i.e.} do \emph{not} decay) but should exhibit a modified \emph{subluminal} refractive index with Lorentz-violating corrections that grow linearly with $E/(M_{\rm QG\gamma}c^2)$, where $M_{\rm QG\gamma}$ is close to the Planck scale,

\item{(ii)} the medium should not refract electrons, so as to avoid the synchrotron-radiation
constraints~\cite{crab,ems}, and

\item{(iii)} the coupling of the photons to the medium must be independent of photon polarization, so as not to have birefringence, thus avoiding the pertinent stringent constraints~\cite{uv,grb,crab2,macio}.

\item{(iv)} The formalism of local effective lagrangians should break down, in the sense that
there are quantum fluctuations in the total energy in particle interactions,
due to the presence of a quantum gravitational `environment', such that stringent constraints, which otherwise would have been imposed from the non-observation of ultra-high energy photons ($\hbar \omega > 10^{19}$ eV), are evaded.
\end{itemize}

 The string-foam models (both type IA and type IIB) are characterized by all these properties~\cite{emnw,ems,emnnewuncert}, and thus avoid the stringent constraints. The absence of birefringence and the transparency  of foam to electrons or charged probes, for reasons of charge conservation, make the models surviving the stringent constraints from synchrotron radiation
 from distant Nebulae~\cite{crab,crab2}.

Moreover, the analysis in \cite{sigl} is based on exact energy momentum conservation in the process (\ref{heirreact}), stemming from the assumption of the local-effective lagrangian formalism for QG-foam. As we discussed in \cite{emngzk,emncomment}, and mentioned in section \ref{sec:uncert}, however, such a formalism is not applicable in the case of the recoiling D-particle space-time foam model, where the fluctuations of space-time or other defects of gravitational nature paly the r\^ole of an external environment, resulting in \emph{energy fluctuations} in the reaction (\ref{heirreact}). The presence of such fluctuations does affect the relevant energy-threshold equations, for the reaction to occur, which stem from kinematics, in such a way that the above stringent limits are no longer valid. In particular, we have seen that the pertinent anomalous dispersion terms are induced by the Finsler-like metric (\ref{opsmetric2}), which in turn arises by the distortion of space time due to the recoil of the D-particle space-time defect during its interaction with the photon. This metric is quadratically suppressed by the string scale, and hence the so-induced modified dispersion relations (\ref{drps}) contain anomalous terms quadratically suppressed by the QG scale, a case for which the constraints coming from ultra high energy cosmic rays are currently weak, but of course the situation can change in future experiments~\cite{sigl}.

It should be stressed again that in the string model the modified dispersion relations are disentangled from the time delays (\ref{timerecoil}) of the more energetic photons, which are linearly suppressed by the string scale. These delays are associated with the string uncertainty principle~\cite{yoneya} and thus are not represented within the local effective field theory framework.
We have also seen that this disentanglement allows for a consistent interpretation of the MAGIC delays with those of the short GRB 090510 observed by FERMI within the string D-foam framework, provided of course
the foam is inhomogeneous.
In addition, as discussed above, the form of the effective string coupling (\ref{effstringcoupl2}), and the induced metric (\ref{opsmetric2}), imply
an upper bound in the photon momentum transfer (\ref{novelgzk}), which characterizes the D-foam model as a result of the sub-luminal nature of the D-particle recoil velocity. These considerations imply that very high energy cosmic rays, when interacting with the D-foam, will have an extremely suppressed interaction rate (the string amplitudes are vanishing when
the recoil velocity approaches the speed of light), thus providing additional reasons~\cite{emncomment} for evading the strong constraints of \cite{sigl}. There are also phenomenologically realistic brane models, with large extra dimensions~\cite{pioline}, for which the ratio $M_s/g_s$ (\emph{i.e.} in our context the D-particle mass) may be of the order of the conventional GZK cutoff~\cite{GZK}, $10^{20}$ eV. In these models, in view of (\ref{novelgzk}), there would be no photons with energies higher than this that could not be completely absorbed when interacting with the D-particle foam. This, therefore, provides an explanation for the absence of such high-energy photons, in accordance with observations, and enables the model to evade the stringent constraints coming from ultra-high-energy cosmic rays.

\section{Conclusions and Outlook \label{sec:5}}
\vspace{0.1cm}
\paragraph{}
In this review we have discussed a stringy version of a (Lorentz-Invariance-Violating) space-time foam model, which
appears as a candidate theory for an explanation of
the delays of the more energetic photons from celestial sources, as observed by MAGIC and FERMI Telescopes, in agreement with all the other current astrophysical tests of Lorentz Violation.
Although such delays could be due to conventional astrophysics at the source, nevertheless
the latter is not well understood at present, and in fact there is no consensus among the relevant communities.

The fact that  string theory (or better its modern version involving D-brane defects) is capable of explaining potentially the observed photon delays, in agreement with all the other astrophysical data currently available,
is at least amusing. This means, if nothing else, that at least such tests provide a framework for experimentally testing/falsifying some models of string theory (entailing Lorentz Violation) at present or in the foreseeable future.
 The key point in the approach is the existence of space-time defects in the ground state of the model, whose topologically non-trivial interactions with the
 string states, via string-stretching during the capture process (\emph{c.f.} fig.~\ref{fig:restoring}),
 are mainly responsible for the observed delays. The latter are found proportional to the incident photon energy.
The situation may be graphically represented by the case of two boats, of different lengths, in rough seas: the waves in the sea represent the foam, the ``medium'' over which the two boats propagate; the longer boat, representing the longer wave-length (lower energy) photons, will cross the waves faster than the smaller boat, corresponding to the higher energy (shorter wavelength) photons, which will be relatively delayed by having to climb up and down the waves.

The peculiarity of the D-particle foam in being \emph{transparent} to charged particles (as a result of electric charge conservation requirements), evades the stringent constraints on linear Planck scale suppression refractive indices that would otherwise have been induced by electron synchrotron radiation studies from Crab Nebula~\cite{crab,crab2}. Moreover, the absence of birefringence avoids the similarly stringent constraints on such models that would have been  imposed by galactic~\cite{crab2} or extra-galactic measurements~\cite{uv,grb}.
Finally, it worths mentioning that the linear in energy time delays (\ref{delayonecapture}),(\ref{redshift}), when applied to neutrinos, can be flavour (\emph{i.e.} neutrino-species) independent, thus avoiding~\cite{ellis08} the stringent constraints that would be obtained from models of quantum gravity with flavour-dependent modifications of neutrino propagation and thus modifications in their oscillations~\cite{brustein}.

We must stress once more, however, that despite the apparent ``success'' of the D-foam model in dealing with
 stringent astrophysical constraints so far, one cannot as yet draw any safe conclusions on the validity of the model in Nature.
 Many more high-energy astrophysical photon measurements are needed in order to disentangle source from possible propagation effects due to fundamental physics.
 If a statistically significant population of data on photons from cosmic sources is collected, exhibiting refractive indices varying linearly with the distance of the source~\cite{mitsou}, as well as the photon energy, then this would be a very strong confirmation of the D-particle foam model, for reasons explained above. However, it must be noted that GRB's, which are expected to lead to statistically significant data in the next few years, will produce photons much lower in energies than the flares observed in AGN, and this could be a drawback. At any rate there are attempts to claim that observations from FERMI will have sensitivity close to the Planck scale~\cite{lamon} for such linear-suppression models soon. The case of GRB090510 is a perfect example of how a single measurement of a short, intense high energy burst, can place stringent limits on Lorentz Violation at the Planck scale. However, caution should be exercised here as to what one means by ``sensitivity at the Planck scale''. As demonstrated above (\emph{c.f}. (\ref{effqgscale2})), the effective quantum gravity scale is actually a complicated function of many microscopic parameters in the model.
 Nevertheless, if many observations at various redshift regimes on delayed arrivals of cosmic photons become available in the future, then we shall be able to make some definite conclusions regarding the order of magnitude of possible quantum gravity effects and thus falsify models, such as the D-particle space-time foam.

We believe we are entering exciting times for tests of fundamental physics concepts and symmetries by means of high-energy astrophysics. Therefore it seems that this branch of physics, together with collider and particle physics, will provide a platform for exploring our world at increasingly smaller microscopic length scales, thereby offering experimental guidance in our quest for the elusive theory of Quantum Gravity.

\section*{Acknowledgements}
\vspace{0.1cm}
\paragraph{}
I would like to thank the organizers of the International Workshop \emph{Recent Developments in Gravity (NEB 14)}, June 8-11 2010, Ioannina (Greece), for their invitation to present this review as a keynote lecture.
This work is partially supported by the European Union
through the Marie Curie Research and Training Network \emph{UniverseNet}
(MRTN-2006-035863).

\end{document}